\documentstyle[prl,aps,12pt]{revtex}

\title{Stress Dependence of Exciton Relaxation Processes in Cu$_2$O}

\author{S. Denev and D.W. Snoke}
\address{Department of Physics and Astronomy, University of Pittsburgh\\
3941 O'Hara St.\\
Pittsburgh, PA 15260}

\begin{document}
\draft
\maketitle
\vspace{1cm}

\begin{abstract}
A comprehensive study of the exciton relaxation processes in Cu$_2$O 
has led to some
surprises.  We find that the ortho-para conversion rate becomes 
slower at high stress,
and that the Auger nonradiative recombination rate increases with 
stress, with apparently
no
Auger recombination at zero stress. These results have important 
consequences for the
pursuit
of Bose-Einstein condensation of excitons in a harmonic potential.
\end{abstract}

\newpage

\section{Introduction}

Cu$_2$O has long been studied as an textbook example of exciton 
physics. In the past
decade, it has attracted attention as a system in which Bose 
condensation of excitons
may occur \cite{lin,mysyr,science,goto}. The interpretation of these 
experiments has
been the
subject of debate, however, because many of the basic relaxation 
processes of the
excitons in
Cu$_2$O have remained unknown.

Two processes in particular have remained controversial. The first is the
Auger nonradiative recombination process. Some authors have argued that Auger
recombination of the excitons effectively prevents Bose condensation 
\cite{ohara}, while
others
have argued that this process is unimportant \cite{kav-mysyr}. The 
theoretical estimates
\cite{kav-baym} of this effect are orders of magnitude different from 
experimental
numbers
\cite{ohara2,auger} .

The second process under debate is the conversion mechanism from orthoexcitons
to paraexcitons. Several authors have argued for a collisional 
conversion mechanism
\cite{kav-mysyr}, while others have argued for a phonon emission mechanism
\cite{ohara}.
Various phonon emission mechanisms have been proposed 
\cite{caswell,snoke-trauer};
previous work indicated a single acoustic phonon emission process 
\cite{snoke-trauer},
but questions have remained.

In this work we present the results of a series of experiments studying these
effects. We have confined the excitons in a harmonic potential to 
keep their volume
nearly constant, and have varied the stress, temperature, and exciton 
density, while
observing both the orthoexciton and paraexciton luminescence as a 
function of time.
This has allowed us to make substantial progress in understanding 
these mechanisms.

\section{Experiment}

For these experiments, we created a harmonic potential trap for the 
excitons using
inhomogeneous stress using the method discussed in a previous 
publication \cite{auger}.
We
created orthoexcitons with kinetic energy approximately 1 meV by 
tuning the laser light
to
photon energy 15 meV above the orthoexciton ground state in the trap.  A strong
phonon-assisted absorption process exists in Cu$_2$O which leads to 
the emission of a
14-meV optical phonon and the creation of an orthoexciton. Because 
the energy of the
excitonic absortion band in the trap is below that of the unstressed 
bulk of the
crystal, when the laser photon energy is tuned to this energy, the 
laser photons
pass through the bulk of the crystal and are only absorbed in the trap.

The laser pulse, generated by a modelocked, cavity-dumped picosecond dye laser,
is 2-5 picoseconds in duration and has negligible uncertainty 
broadening, much less
than 1 meV. The cavity dumper allows us to have a long period between pulses
(typically 260 ns) so that almost no excitons remain from one 
excitation pulse to the
next. We observe the luminescence from the paraexcitons and 
orthoexcitons in the trap
as a function of time after the laser pulse. Orthoexcitons decay 
either via a single-photon
direct emission process or a phonon-assisted recombination, while 
paraexcitons decay
primarily only via a single-photon direct recombination process, 
which is forbidden by
symmetry at zero stress but which becomes allowed at high stress. The
orthoexciton phonon-assisted luminescence intensity is directly 
proportional to the total
orthoexciton population at all times; the paraexciton 
direct-recombination luminescence
intensity is proportional to the paraexciton population and also, in 
principle, depends on
the
temperature, since only low-momentum paraexcitons can participate in the direct
recombination
process \cite{snoke-short}. Our fits indicate that this temperature 
dependence is very
weak,
however, and therefore we simply assume that the paraexciton 
luminescence intensity is
directly
proportional to the paraexciton density in fitting the data.

We can measure the energy spectrum of the exciton luminescence with 40 ps time
resolution using time-correlated single photon counting. It is 
important to use the entire
time-resolved spectrum of the luminescence for these measurements, 
because incorrect
conclusions can be deduced by looking only at luminescence from one 
wavelength.  We
can also
measure the spatial profile of the luminescence with the same time 
resolution using the
scanner method reported elsewhere \cite{scanner}.

The stresses are calibrated using the luminescence line positions of 
the orthoexcitons
and paraexcitons, based on the reported shifts of the lines with stress
\cite{trauer,waters}.

The crystal temperature was held at 2 K in immersion in liquid helium 
for most of the
experiments; we also performed experiments at higher temperature in 
helium vapor. The
excitons
have higher than the lattice temperature for a few nanoseconds after 
the laser pulse, since
they are created with 1 meV excess energy, and the orthoexciton 
generation process also
creates
optical phonons which may cause a local heating.

\section{Stress Dependence of the Ortho-Para Conversion Rate}

The first surprise from these experiments is that the ortho-para 
conversion rate becomes
slower with increasing stress. Fig. 1 shows orthoexciton luminescence 
decay data for
several stresses at very low excitation density. The decay is single 
exponential in
each case. Because there are also density-dependent processes in 
Cu$_2$O, as discussed
in the Section \ref{augersect}, we used very low laser power (1 mW) for these
measurements. We
verified that the time evolution did not depend on laser power by recording the
orthoexciton lifetime for two powers different by a factor of ten.

Fig. 2 shows a summary of the orthoexciton decay time at low density 
as a function of
the ortho-para splitting energy, which in turn depends on the stress. 
In general, the
decay of the orthoexcitons at low density comes not only from 
ortho-para conversion but
also from radiative recombination of orthoexcitons and recombination 
at impurities. As
shown in previous measurements at high temperature 
\cite{snoke-shields} (verified by
this
study, as discussed in Section \ref{surfacesect}), the orthoexciton radiative
recombination
lifetime is greater than 300 ns, so that radiative recombination of 
orthoexcitons gives
negligible contribution to the orthoexciton decay.

We can understand this effect in terms of emission of a single 
acoustic phonon, as
illustrated
in the inset of Fig. 2. This mechanism was proposed in a previous publication
\cite{snoke-trauer} to explain the temperature dependence of the 
luminescence lines in
Cu$_2$O. That publication successfully explained some results, but 
contained an error in
the
$k$-dependence. Correcting that error allows us to fit the data with 
the curve shown in
Fig. 2.

The dependence on $\Delta$ can be understood as arising from a 
$k$-dependence of
the acoustic phonon emission process. As the ortho-para splitting is 
decreased, the
momentum of the phonon emitted is decreased, as illustrated in the 
inset of Fig. 2. Since
at
low temperature the orthoexcitons have nearly zero $k$-vector, and 
since the acoustic
phonon
emission is essentially a horizontal process, the $k$-vector of the phonon is
essentally equal to the momentum of the final para state. Since the 
paraexciton kinetic
energy
is proportional to $k^2$, the magnitude of $k$ is proportional to 
$\Delta^{1/2}$.

In Ref. \cite{snoke-trauer}, the ${\bf k} \cdot {\bf p}$ matrix 
element was given as
proportional to
$k_o$, the orthoexciton momentum, but this is an incorrect estimation 
of the matrix
element. In ${\bf k} \cdot {\bf p}$ theory \cite{ridley}, the 
orthoexciton and paraexciton
states can be written as
\begin{eqnarray}
| o ({\bf k}) \rangle &=& |o(0)\rangle + | i\rangle \ {\bf k}_o\cdot 
\langle i | {\bf p}
| o(0) \rangle
\nonumber \\
| p ({\bf k}) \rangle &=& |p(0)\rangle + | i \rangle \ {\bf k}_p\cdot 
\langle i | {\bf
p} | p(0) \rangle
\nonumber
\end{eqnarray}
where ${\bf k}_o$ and ${\bf k}_p$ are the ortho and para momenta, 
respectively, and $| i
\rangle$ is some intermediate state. The phonon scattering rate is 
therefore proportional
to
\begin{equation}
|\langle o ({\bf k}_o)| M_{\bf q} | p ({\bf k}_p)\rangle|^2 = 
\left|\langle o(0) | M_{\bf q}
|i\rangle  {\bf k}_p\cdot \langle i | {\bf p} | p(0) \rangle + 
\langle i | M_{\bf q} | p(0)
\rangle  \langle o(0) |{\bf p}| i \rangle\cdot {\bf k}_o
  \right|^2
\end{equation}
where $M_{\bf q}$ is the exciton-phonon deformation potential matrix 
element for a
phonon with
momentum ${\bf q}$.  At low temperature, the orthoexciton
momentum ${\bf k}_o$ is negligible compared to the paraexciton 
momentum. Therefore
we can drop
the second term above, and assume that the rate is proportional to
$k_p^2$ and set ${\bf q} = {\bf k}_p$. The deformation potential 
matrix element for
phonon
emission is given by \cite{seeger}
\begin{equation}
|M_{\bf q}|^2 = \frac{\hbar \Xi^2 q}{2\rho v V}(1+n_{\bf q})
\end{equation}
where $\Xi$ is a deformation potential, $\rho$ is the crystal mass 
density, $v$ is the
acoustic phonon velocity, and $n_{\bf q}$ is the number of phonons in 
state ${\bf q}$.
Since
this process involves spin conversion, the deformation potential 
$\Xi$ used here is
not necesarily the same as the deformation potential for acoustic 
phonon relaxation
wihout spin flip, determined in previous experiments \cite{snokebc}. 
For an orthoexciton
with
$k_o
\simeq 0$, the phonon emission rate is therefore proportional to
\begin{eqnarray}
\Gamma &\propto& \int d^3k_p \ \left[ k_p^3 
\Xi^2\delta(\Delta-\hbar^2k_p^2/2m -
\hbar
vk_p)\left(1+\frac{1}{e^{\hbar vk_p/k_BT}-1} \right)\right. \nonumber\\
&& + \left.\ k_p^3 \Xi^2 \delta(\Delta-\hbar^2k_p^2/2m + \hbar
vk_p)\left(\frac{1}{e^{\hbar
vk_p/k_BT}-1}\right)  \right],
\label{opconv}
\end{eqnarray}
where we have included both the phonon emission and absorption terms. 
The phonon
absorption
term is similar. Using the $\delta$-function to remove the 
integration over $k_p$
yields
\begin{equation}
\Gamma \propto \frac{k_p^5}{\hbar k_p /m + 
v}\left(1+\frac{1}{e^{\hbar vk_p/k_BT}-
1}\right)
+ \frac{k_p'^5}{\hbar k_p' /m - v}\left(\frac{1}{e^{\hbar 
vk_p'/k_BT}-1}\right),
\end{equation}
where $k_p = (\sqrt{2\Delta m} - vm)/\hbar$ and $k_p' = (\sqrt{2\Delta m} +
vm)/\hbar$. This formula gives the curve
shown in Fig. 2, which is fit to the data by an overall multiplier.

Ref. \cite{snoke-trauer} reported essentially no stress dependence of 
the ortho-para
conversion rate, but the data were over a much narrower range of 
stress, and the
conversion
rate was deduced based on the total intensities, which is a much more 
indirect method
than
that used here.

The phonon-assisted conversion rate (\ref{opconv}) also implies that 
the ortho-para
conversion
rate depends on temperature. In the limit of low temperature, phonon 
absorption is
negligible, and $n_{\bf q} \simeq k_BT/vK_p$, which implies $\Gamma 
\propto (1 +
aT)$. A weak
temperature dependence consistent with this prediction has been 
reported in Ref.
\cite{weiner}
for the case of zero stress; we also have measured a weak temperature 
dependence
consistent
with this formula; the ortho-para conversion rate increases by 
approximately a factor of
two
when the temperature is raised from 2 K to 10 K.  This is one of the 
strong arguments
against
an early proposal \cite{caswell} that an optical phonon is involved 
in the conversion
process;
if an optical phonon was emitted, the rate would increase much more 
rapidly with
temperature.

\section{Stress Dependence of the Auger Constant}
\label{augersect}

It has long been proposed that an Auger process exists in Cu$_2$O by which two
excitons
collide, resulting in the annihilation of one and ionization of the 
other. Strong evidence of
this process was provided by strain well experiments by Trauernicht 
et al. \cite{trauer-
well}.

Ref. \cite{auger} reported the measurement of an Auger rate constant 
for excitons in
a strain well in Cu$_2$O.  That work reported data for only one, 
moderate stress (2.5
kbar).
Furthermore, at early times after the laser pulse, the paraexciton 
data were obscured by
hot
orthoexciton luminescence. In the present work, we report data for 
several stresses, and
we
have clearly resolved the paraexciton luminescence by recording the 
full luminescence
spectrum at all times after the laser pulse for all excitation densities.

Surprisingly, we find that the Auger rate depends on the stress. This 
is consistent with the
prediction of Baym and Kavoulakis \cite{kav-baym} that the Auger rate should be
negligible at
zero stress and increase with stress. We see no evidence of a 
two-body spin-flip
mechanism
\cite{kav-mysyr}, however.

Figs. 3 and 4 show the integrated ortho and paraexciton luminescence 
intensities for two
different laser powers, in a strain well with 3.5 kbar stress, at $T 
= 2$ K. The evolution is
clearly different at high exciton density. The dark lines are a fit 
to the coupled rate
equations
\begin{eqnarray}
\frac{dn_o}{dt} &=& -\frac{n_o}{\tau_{o-p}} - A_on_o^2 + \frac{3}{8}(A_on_o^2 +
A_pn_p^2)
-\frac{n_o}{\tau_o}\nonumber
\\
\frac{dn_p}{dt} &=& \frac{n_o}{\tau_{o-p}} - A_pn_p^2 + \frac{1}{8}(A_on_o^2 +
A_pn_p^2)
-\frac{n_p}{\tau_p}.
\label{rate}
\end{eqnarray}

The terms with $A_o$ and $A_p$ are the ortho and para Auger rates, 
respectively. One
half of
these terms are added back to the ortho and para populations due to 
ionized excitons
reforming; it is assumed that 3/4 of these return as orthoexcitons 
and 1/4 as paraexcitons
since the spin is assumed to be randomized upon ionization. In 
general, we could also
include
a cross term for collisions between orthoexcitons and paraexcitons, 
but we have been
unable to
fit the data with such a term.

The ortho-para conversion time $\tau_{o-p}$ depends on temperature, 
as indicated by
Eq.
(\ref{opconv}); as shown in Ref. \cite{caswell} and confirmed by this 
study, the
conversion
rate is faster at higher temperature.  Since the temperature falls in 
time, the ortho-para
conversion rate  is assumed to vary in time according to
\begin{equation}
\frac{1}{\tau_{o-p}(t)} = \frac{C\exp(-t/\tau_T) + 1}{\tau_{o-p}},
\label{timedep}
\end{equation}
where $\tau_T$ is on the order of 10 ns and $C$ is a unitless 
constant which gives the
fractional increase of the rate.  These are not free parameters, because
measurements of the spectral width of the exciton luminescence as a 
function of time
constrain
the temperature evolution.  Fig. 5 shows the full width at half 
maximum (FWHM) of the
orthoexciton phonon-assisted luminescence for the two cases of Figs. 
3 and 4. The dark
lines
are the FWHM implied by the fits of Figs. 3 and 4, because the FWHM 
of the phonon-
assisted line
in a three-dimensional harmonic potential is equal to $3.4k_BT$. At 
late times, both lines
have
FWHM of 0.75 meV, which corresponds to the expected FWHM of the phonon-assisted
line at $T=2$ K
convolved with the spectral resolution of 0.5 meV.  (Our spectral linewidth
measurements also
confirm the result reported by Trauernicht et al. \cite{trauer-well}, 
that the paraexciton
luminescence in the strain well always has narrower spectral width 
than the orthoexciton
luminescence at the same temperature, due to the peculiarities of the 
stress-allowed
direct recombination process.)

The lifetimes $\tau_o$ and $\tau_p$ represent the total recombination 
rate due to
radiative
recombination and recombination at impurities. As shown earlier, and 
confirmed here,
the
radiative recombination rate for orthoexcitons and paraexcitons in 
Cu$_2$O is extremely
low
and essentially negligible. We find that in this sample the 
paraexcitons have a total
recombination rate on the order of 20 ns at low temperature, however. 
This is presumably
due
to recombination at impurities. These fits indicate that the conversion to
paraexcitons is the dominant decay channel for orthoexcitons, 
however, which means that
their
lifetime for recombination at impurities must be greater than 100 ns. 
If the orthoexcitons
decayed by other channels than conversion to paraexcitons, we would 
not see as many
paraexcitons at early times.

The paraexciton luminescence efficiency is different from that of the 
orthoexcitons;
therefore
another parameter is introduced in these fits which is an overall 
multiplier for the
paraexciton luminescence. This overall multiplier depends on the 
stress since the
paraexciton
radiative recombination rate increases as the square of the shear 
stress. As mentioned
above,
in principle this multiplier also depends on temperature and 
therefore also changes in
time,
but we do not see evidence for this effect. One reason may be that 
the polariton region of
the
paraexcitons covers a spectral region comparable to $k_BT$ at high stress.

Although these fits have several parameters ($\tau_{o-p}, \tau_p, 
A_o, A_p$, and the
overall
multiplier for the para intensity, $R_p$), the fits are highly 
constrained by the large
amount
of data. The same set of parameters must fit both the ortho and para 
data at all times and
at
all densities, with the exception that we allow the constant $C$ used 
in the time
dependence
of $\tau_{o-p}$ to become higher at high density, consistent with the 
increase of
temperature
which occurs when the Auger process is important.

The same set of equations have been used to fit the ortho and para 
data at other stresses.
Figs. 6 and 7 show the fits for 2.5 kbar.  As seen in these figures, 
the fits to Equations
(\ref{rate}) are very good. Fig. 8 shows the fit for high exciton 
density at 1.5 kbar. At this
stress the orthoexciton luminescence partially obscures the 
paraexciton data at early
times, so
that we cannot fit the paraexciton data at those times. 
Surprisingly, however, for the 1.5
kbar orthoexciton data there is essentially no Auger effect, implying 
that the Auger rate
depends on the stress.  Fig. 9 shows the orthoexciton data at 1.5 
kbar at three laser
powers,
100 mW, 10 mW, and 1 mW, multiplied by overall factors of 1, 10, and 100,
respectively. As
seen in this figure, all the curves decay in the same way; there is 
no evidence for any
density
dependence. The nonexponential nature of the decay can be explained 
as due to the
temperature
dependence of the ortho-para conversion process. Data taken at 2 kbar 
show similar
behavior,
with only slight dependence of the decay on the exciton density. We 
can be certain that
the
exciton densities are comparable at the different stresses, because 
the number of photon
counts from the phonon-assisted orthoexciton luminescence immediately 
after the laser
pulse is
approximately the same in each case, and the ortho phonon-assisted 
recombination rate
does not
depend on stress.

Table 1 shows the rate parameters deduced from these fits at 2 K.  The standard
deviations of
the parameters are generally small because each set of parameters 
represents a single fit
to
six data sets, the ortho and para data at three different densities. The
uncertainty in the measurement of the stress is approximately 0.1 
kbar.  Since the exciton
volume does not depend strongly on the applied stress, this study 
indicates that the Auger
rate
increases roughly as the square of the stress. The relative values of the
$A_o$ and $A_p$ for different stresses are tightly constrained by 
these fits, but the
absolute
values are uncertain within a factor of two because of the 
uncertainty in the absolute
density
of the excitons. Setting the absolute magnitudes of the Auger 
coefficients requires an
estimate
of the absolute exciton density in the strain well, which is done by 
the same process as
previously reported \cite{auger}.  The present results are consistent 
with a previous study
\cite{auger} which gave a value of approximately $10^{-16}$ cm$^3$/ns 
for both $A_o$
and $A_p$
at 2.5 kbar. The average relative number of 0.001 for the Auger 
constants at 2.5 kbar
should
therefore be equated with this absolute value.

As mentioned above, some authors \cite{kav-mysyr,other} have proposed that two
orthoexcitons
can convert rapidly to two paraexcitons by a collisional process. We 
cannot strictly rule
out
this process, but find it is not necessary to invoke this process to 
fit the data. It is not
possible to fit the data at high stress without an Auger process. The 
Auger process shows
up
in two important ways. First, the paraexciton rise time is shorter at 
high density than at
low
density, as shown in Figs. 3(b) and 6(b). This happens because as the
paraexciton density increases, the inflow is counteracted by 
increasing outflow due to
Auger
recombination. A second effect is that extra orthoexcitons are 
generated by up-conversion
from
paraexcitons, which leads to a slowing of the ortho decay at late 
times at high density.
This
effect of up-conversion from paraexcitons to orthoexcitons in the 
strain well has been
reported before \cite{trauer-well}.

In general, it is not possible to prove that no other processes 
besides those included in the
rate equations (\ref{rate}) can exist. We can only say that these 
equations fit the data well
over a wide range of density, temperature and stress, and this lends 
strong support to the
existence of a stress-dependent Auger effect.

\section{Surface Effects?}
\label{surfacesect}

The fact that we do not see evidence for an Auger effect at 1.5 kbar 
is surprising, since
previous authors have reported evidence for an Auger effect at zero stress
\cite{ohara2,warren}. One possible difference is that those 
experiments involved
excitation of
the surface of the crystal, using light absorbed within 25 microns of 
the surface.  The
presence of the surface may change the local symmetry enough to cause 
an Auger effect.

We have verified that an Auger effect must occur in surface 
excitation with zero stress.
Fig.
10 shows the orthoexciton luminescence intensity for high temperature (300 K)
excitation, with
negligible applied stress. The absorption length of the 600 nm laser 
at this temperature is
approximately 10 $\mu$m \cite{paster}, which effectively confines the 
excitons to a
region
near the surface, as in the case of 514 nm excitation at low 
temperature. At room
temperature,
the orthoexcitons and paraexcitons are strongly coupled by phonon 
absorption and
emission;
therefore the orthoexciton decay gives the total decay of the two 
species of excitons. As
seen
in this figure, the excitons decay more rapidly at high density, 
indicating a density-
dependent
process as expected for an Auger process. A spin-flip mechanism as 
suggested by Ref.
\cite{kav-mysyr} would not give this behavior, since it conserves the 
total number of
excitons. Yet at low stress in the well, we did not see any evidence 
for an Auger process.
One
possible explanation is that the density is much higher in the case 
of surface excitation.
We
do not have any direct measurement of the depth of the region of the crystal
excited by the laser, but estimating a depth of 10 $\mu$m based on the reported
absorption
constant, and a laser spot diameter of 200 $\mu$m, the density of the 
excitons in this case
should be lower than in the strain well trap measurements discussed 
above. Another
possible
explanation is that the surface plays a role in the Auger effect. 
Many processes, such as
second harmonic generation, are forbidden in the bulk of a 
centrosymmetric crystal like
Cu$_2$O but are allowed near a surface.

At late times the orthoexciton luminescence fits a single-exponential 
decay with a
lifetime of
350 ns, consistent with earlier results \cite{snoke-shields}.  Since 
the orthoexcitons and
paraexcitons are well coupled, this is also a lower bound for the 
paraexciton nonradiative
decay lifetime. At low temperature, we found a paraexciton decay rate 
of around 20 ns, as
discussed in Section \ref{augersect}, which cannot be the case here. 
Presumably, the
paraexcitons can only become bound to impurities at low temperature, 
and therefore do
not
decay by that mechanism at room temperature.

This 350 ns lifetime does give a lower bound on the radiative 
recombination rate at low
temperature. At higher temperature, the direct single-photon 
recombination process for
orthoexcitons is nearly forbidden, but the phonon-assisted process is 
not strongly affected
by
temperature \cite{snoke-shields}.  The phonon-assisted radiative 
recombination process
cannot
occur with a lifetime less than 350 ns, based on these measurements. 
At low temperature,
the
total direct recombination luminescence intensity is approximately 
the same as that of the
phonon-assisted recombination luminescence. Therefore, even at low 
temperature, the
radiative
recombination lifetime of the orthoexcitons cannot be less than 
around 100 ns. The
paraexciton
phonon-assisted radiative lifetime is known from absorption 
experiments to be 500 times
longer
than the orthoexciton radiative lifetime \cite{para500,snokeeverr}. 
As stress is increased,
the
paraexciton radiative lifetime becomes shorter, but the total 
paraexciton direct
recombination
luminescence intensity does not exceed the orthoexciton direct recombination
luminescence
intensity, which implies that the paraexciton radiative lifetime must 
also be on the order
of
100 ns even at high stress.

\section{Conclusions}

Previous experiments have concentrated on seeking BEC of excitons in 
Cu$_2$O in one
of two
types of experiments-- either with intense surface excitation, or in 
harmonic-potential
traps
made with high stress.  These results indicate that searches for BEC 
may be best pursued
in a
trap with fairly low stress. With surface excitation, surface effects 
may shorten the
lifetime
of the excitons. At high stress, the Auger recombination rate seems 
to increase, and the
orthoexciton conversion to paraexcitons gets slower. Both of these 
effects inhibit Bose
condensation in the exciton ground state.  If excitons are created as 
orthoexcitons, then at
high stress they will remain in that state longer, instead of 
converting into the
lower paraexciton state, where we would like to accumulate the condensate.

The problem with pursuing BEC of excitons at low stress is that the 
orthoexciton
luminescence
line substantially overlaps the paraexciton line in energy, which is 
also weaker since the
paraexciton line is forbidden at zero stress. This makes analysis of 
the spectral lines
difficult. One obvious solution is to not create orthoexcitons. This 
is not so simple,
because
the paraexciton absorption is very weak, and therefore most 
experiments have used the
strong
phonon-assisted orthoexciton absorption. A way around this is to use 
resonant two-
photon
generation of paraexcitons. Experiments with this approach are under way.

{\bf Acknowledgments}.  This work has been supported by the National Science
Foundation as part of Early Career award DMR-97-22239.  We thank V. Negoita for
early
contributions to these experiments, and R. Hurka for contributions to 
the analysis.
Samples of
Cu$_2$O were obtained from P.J. Dunn of the Smithsonian Institute and from J.P.
Wolfe.

\newpage
\begin{table}
\begin{tabular}{|c|l|l|l|}
   & 1.5 kbar & 2.5 kbar & 3.5 kbar \\ \hline
$\tau_{o-p}$ (ns) & 3.2  $\pm 0.1 $ & 4.9 $\pm 0.03$  & 9.4 $\pm 0.06$  \\
$\tau_p$ (ns) & 14.6 $\pm 0.3$ & 15.2 $\pm 0.2$  & 19.4 $\pm 0.3$  \\
$A_o$ (relative units) & 0.00046 $\pm 0.00001$ & 0.0015 $\pm 0.00002$   & 0.003
$\pm
0.00004$
\\
$A_p$ (relative units) & 0.000136 $\pm 0.000005$ & 0.00074 $\pm 
0.00002$ & 0.0038
$\pm .0001$
\\
$R_p$ & 0.0142 $\pm$ 0.0007 & 0.047 $\pm 0.0007$ & 0.186 $\pm 0.002$ \\
\end{tabular}
\caption{Fit values from least-squares fits of the data sets to the 
theory discussed in the
text.}
\end{table}

\newpage

FIGURE CAPTIONS

\noindent FIG 1. Decay of the orthoexciton phonon-assisted 
luminescence at three
different
stresses, under identical excitation conditions.
\vspace{1cm}

\noindent FIG 2. Solid circles: lifetime of the orthoexciton 
phonon-assisted luminescence
at
low density as a function of the ortho-para energy splitting. Open 
circle: lifetime of the
orthoexcitons at zero stress, reported by Weiner et al. 
\protect\cite{weiner}.  Solid line:
fit
to the theory discussed in the text.  Inset: the relative energies of 
the orthoexciton (upper
level), paraexciton (lower level) and acoustic phonons involved in 
the conversion
process.
\vspace{1cm}

\noindent FIG 3. Ortho and para luminescence decay in the strain well 
for the case $T=
2$ K,
3.5 kbar stress, and 100 mW average laser power. (a) Open circles: 
intensity of the
orthoexciton phonon-assisted luminescence as a function of time 
following the laser
pulse.
Solid line: fit to the theory discussed in the text. (b) Open 
circles: intensity of the
paraexciton direct recombination luminescence as a function of time. 
Solid line: fit for
the
same theory as (a).
\vspace{1cm}

\noindent FIG 4. Ortho and para luminescence decay in the strain 
well, as in Fig. 3, but
for
the case $T= 2$ K, 3.5 kbar stress, and 10 mW average laser power.
\vspace{1cm}

\noindent FIG 5. Full width at half maximum (FWHM) of the orthoexciton phonon-
assisted
luminescence spectrum for the same data used in Figs. 3 and 4. Solid 
lines: the FWHM in
each
case implied by the temperature used in the orthoexciton 
down-conversion rate of the fits
to
the data of Figs. 3 and 4.
\vspace{1cm}

\noindent FIG 6. Ortho and para luminescence decay in the strain 
well, as in Fig. 3, but
for
the case $T= 2$ K, 2.5 kbar stress, and 100 mW average laser power.
\vspace{1cm}

\noindent FIG 7. Ortho and para luminescence decay in the strain 
well, as in Fig. 3, but
for
the case $T= 2$ K, 2.5 kbar stress, and 10 mW average laser power. 
(At late times, the
orthoexciton luminescence hits the dark count rate, and this dark count rate
is included in the fit as a additive constant of 0.01.)
\vspace{1cm}

\noindent FIG 8. Ortho and para luminescence decay in the strain 
well, as in Fig. 3, but
for
the case $T= 2$ K, 1.5 kbar stress, and 100 mW average laser power. 
At early times the
orthoexciton luminescence obscures the paraexciton luminescence.
\vspace{1cm}

\noindent FIG 9. Ortho luminescence decay in the strain well for the case
$T= 2$ K, 1.5 kbar stress, and three different laser powers. Circles: 
1 mW. Squares: 10
mW.
Diamonds: 100 mW.
\vspace{1cm}

\noindent FIG 10. Orthoexciton luminescence decay at $\lambda = 629$ 
nm for surface
excitation
at $\lambda = 602$ nm, at room temperature and zero stress, for two 
laser powers. The
two
curves are normalized to the same initial value. The laser focus spot size was
approzimately
200 $\mu$m.

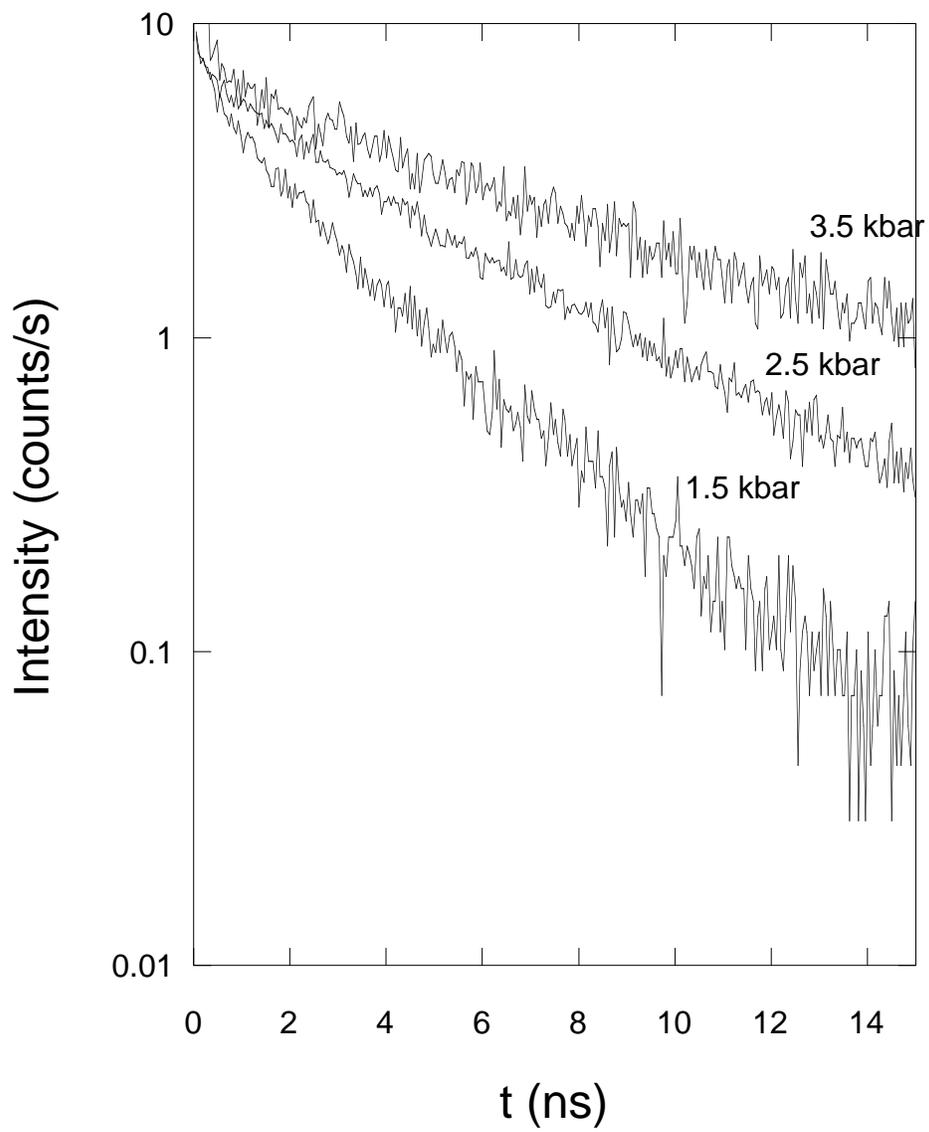
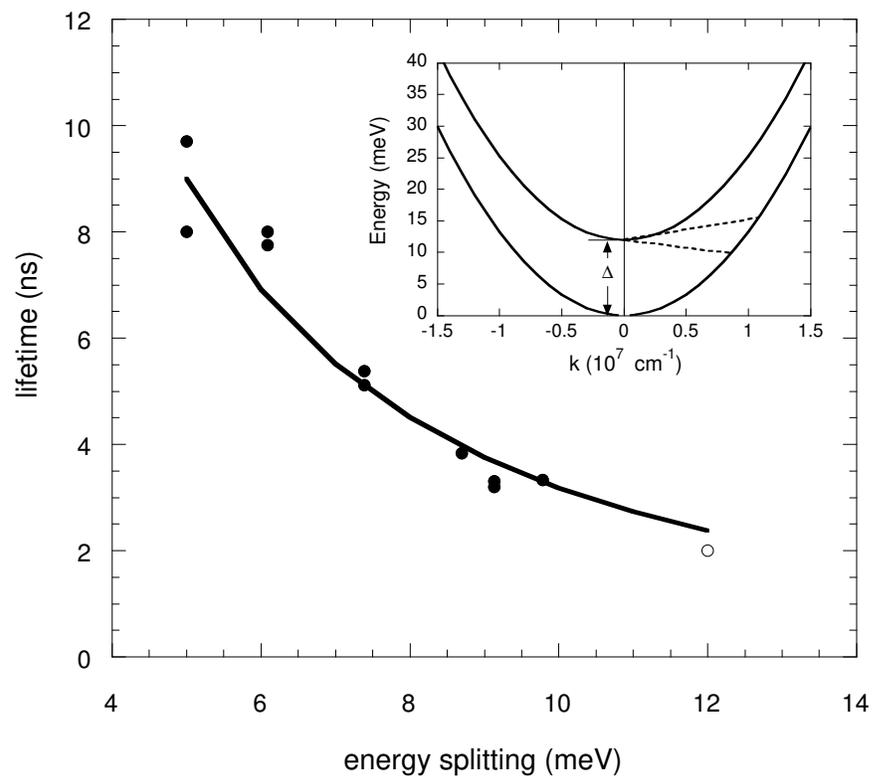
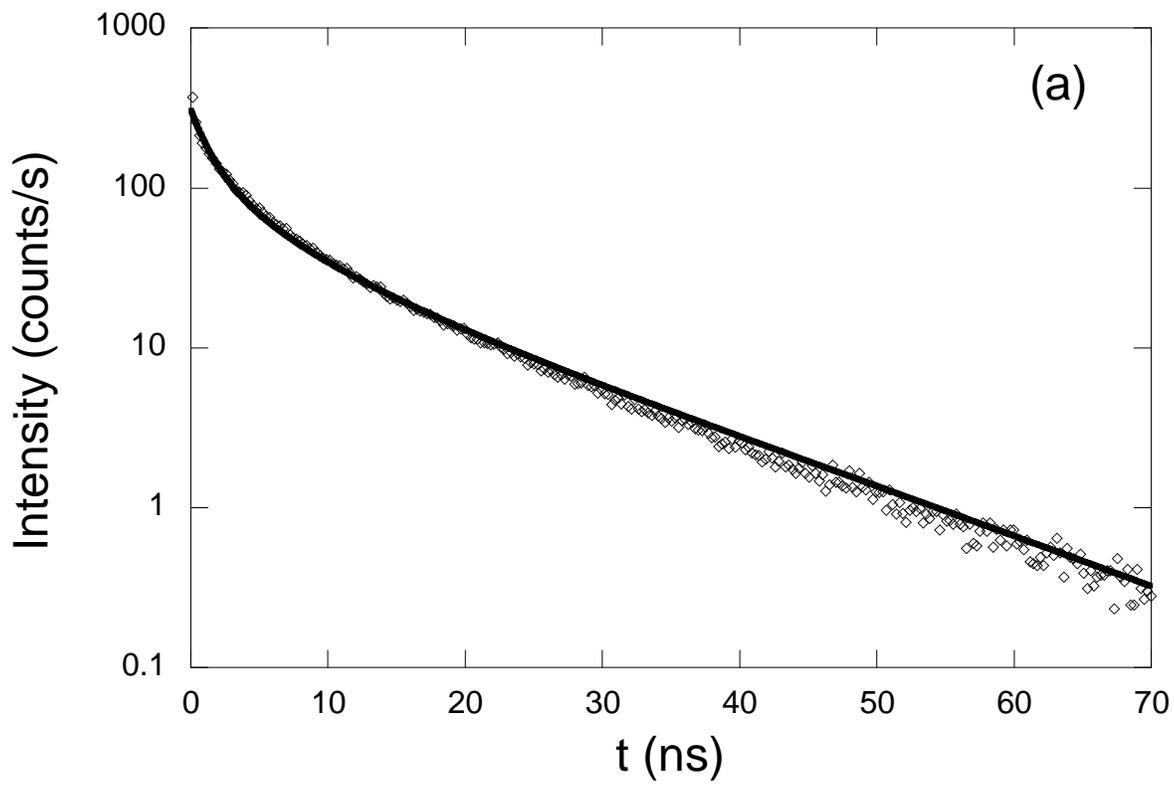
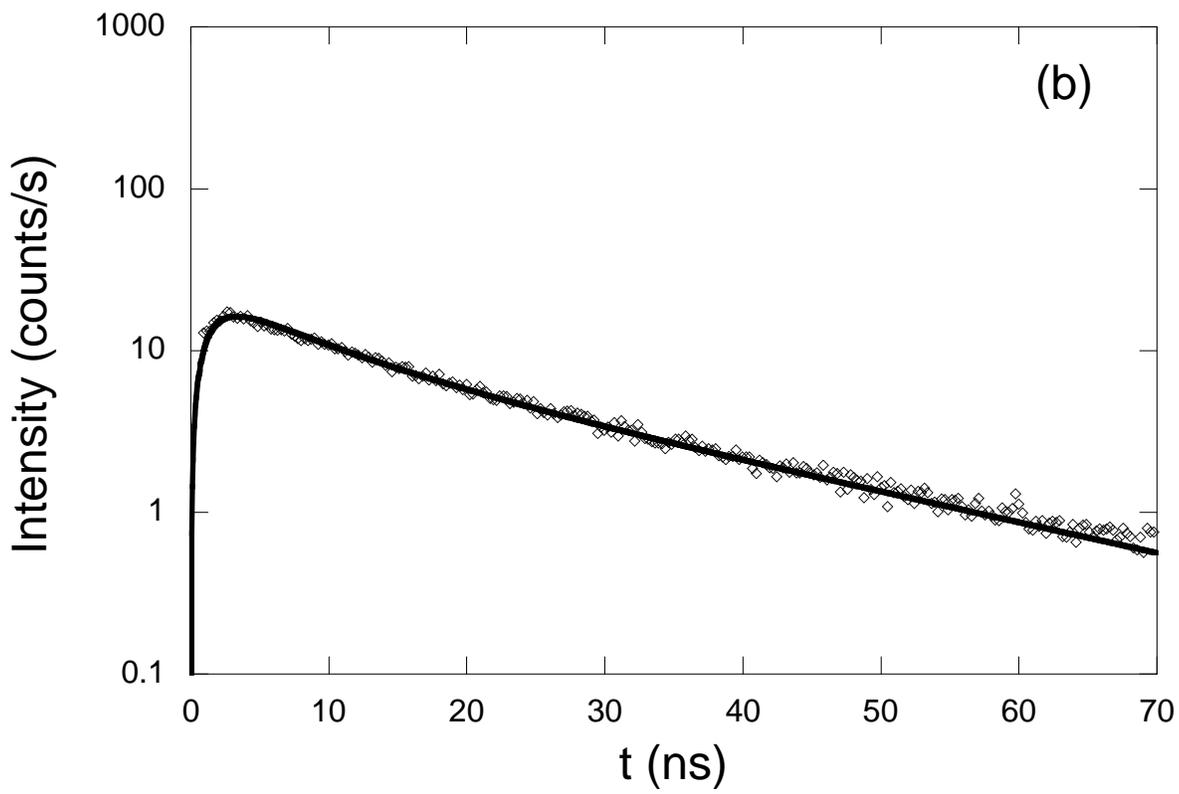
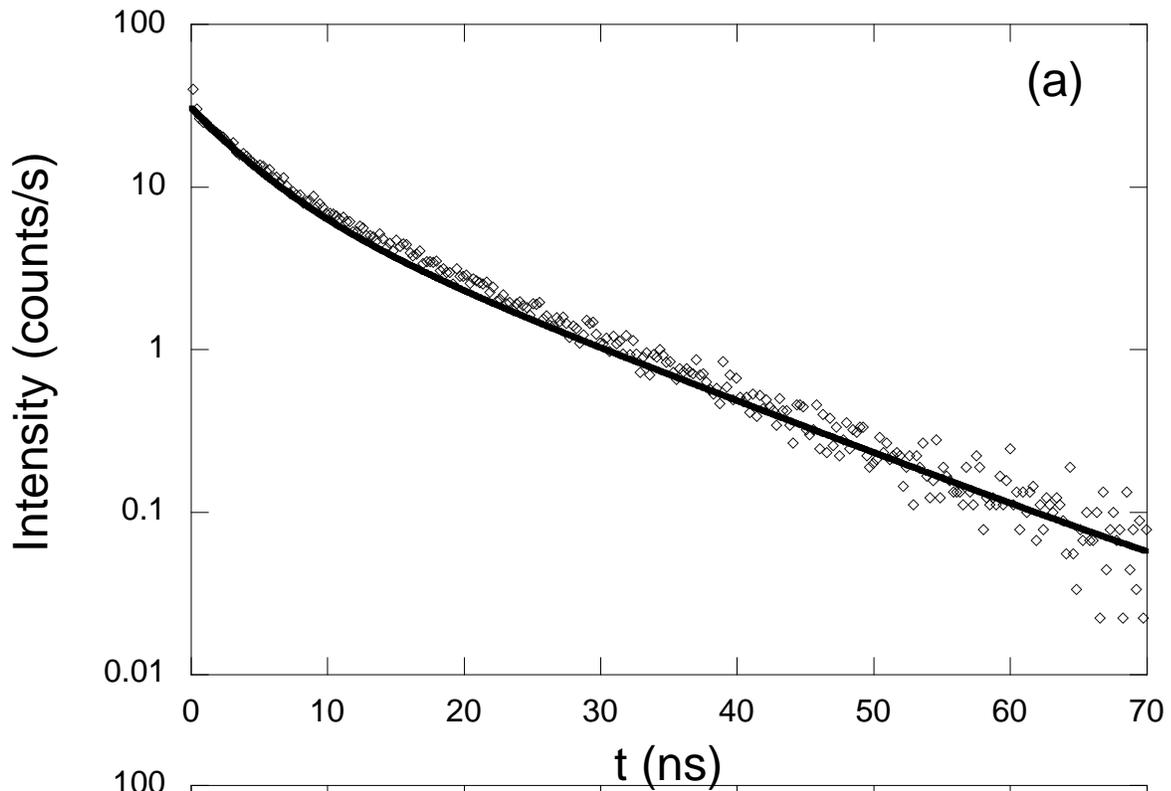
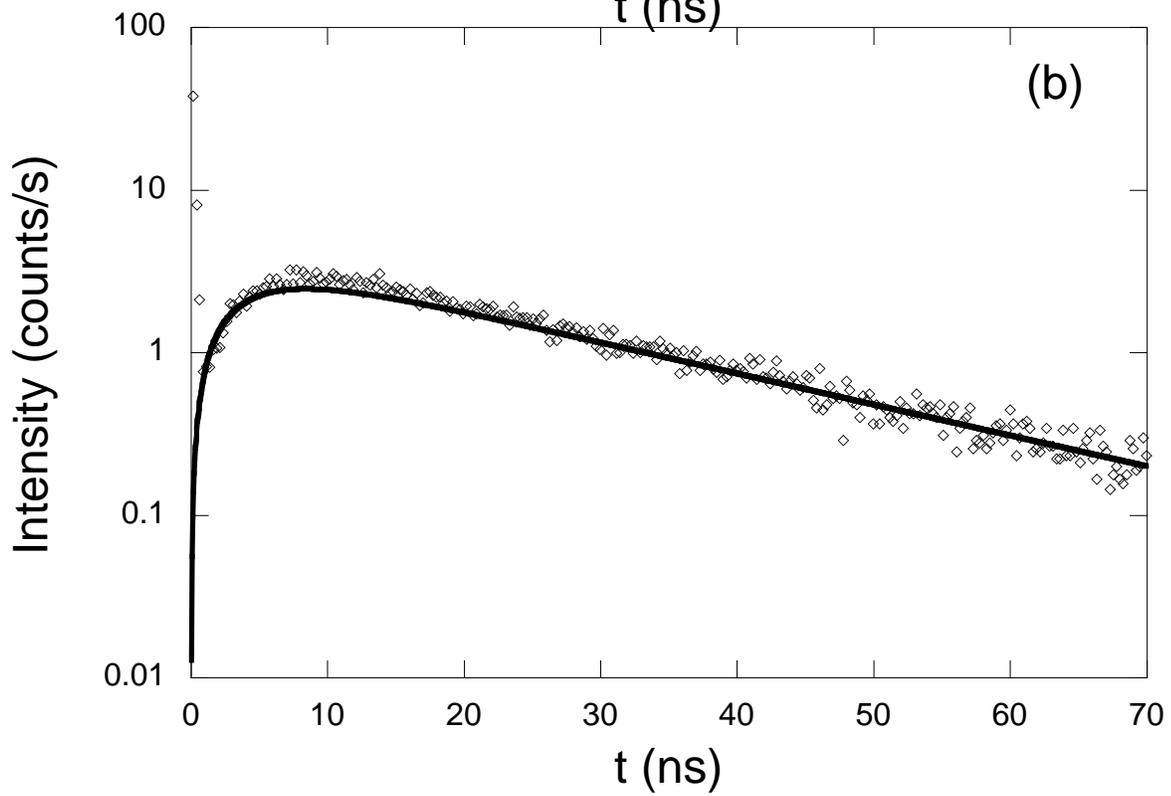
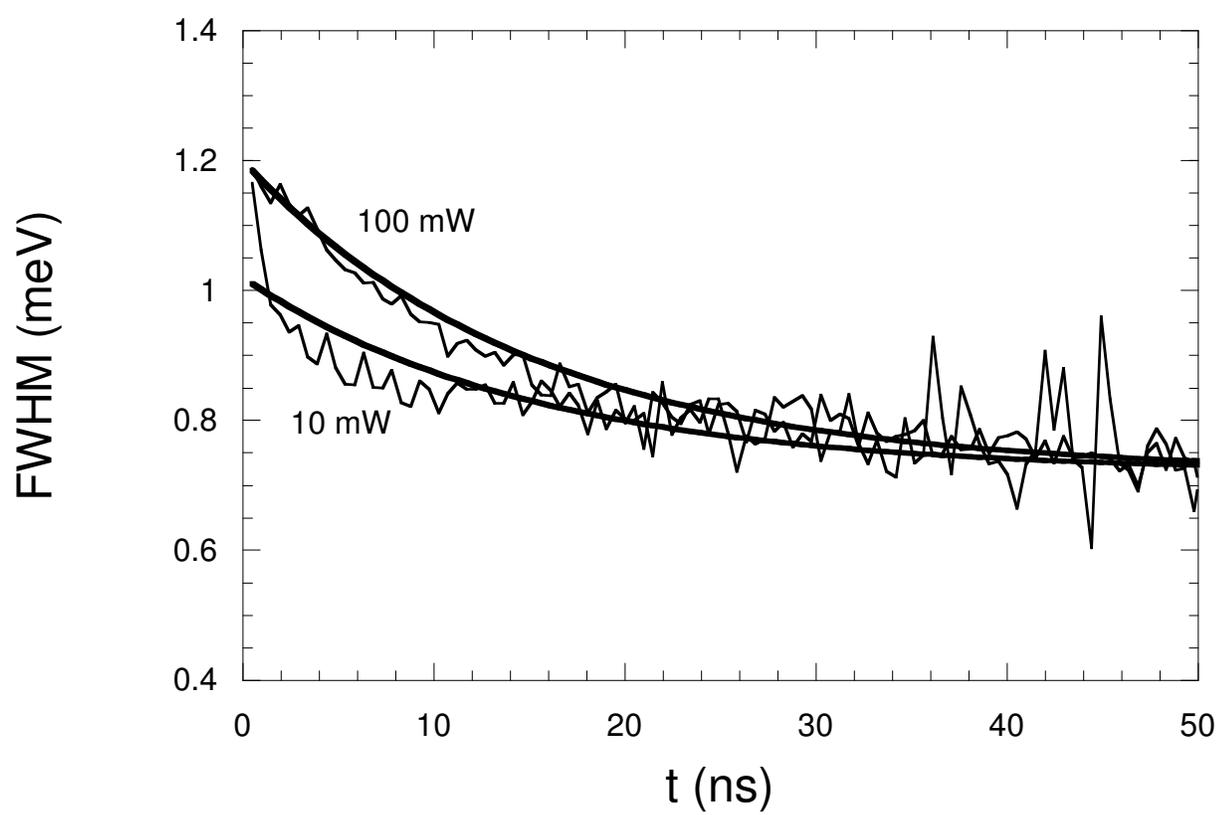
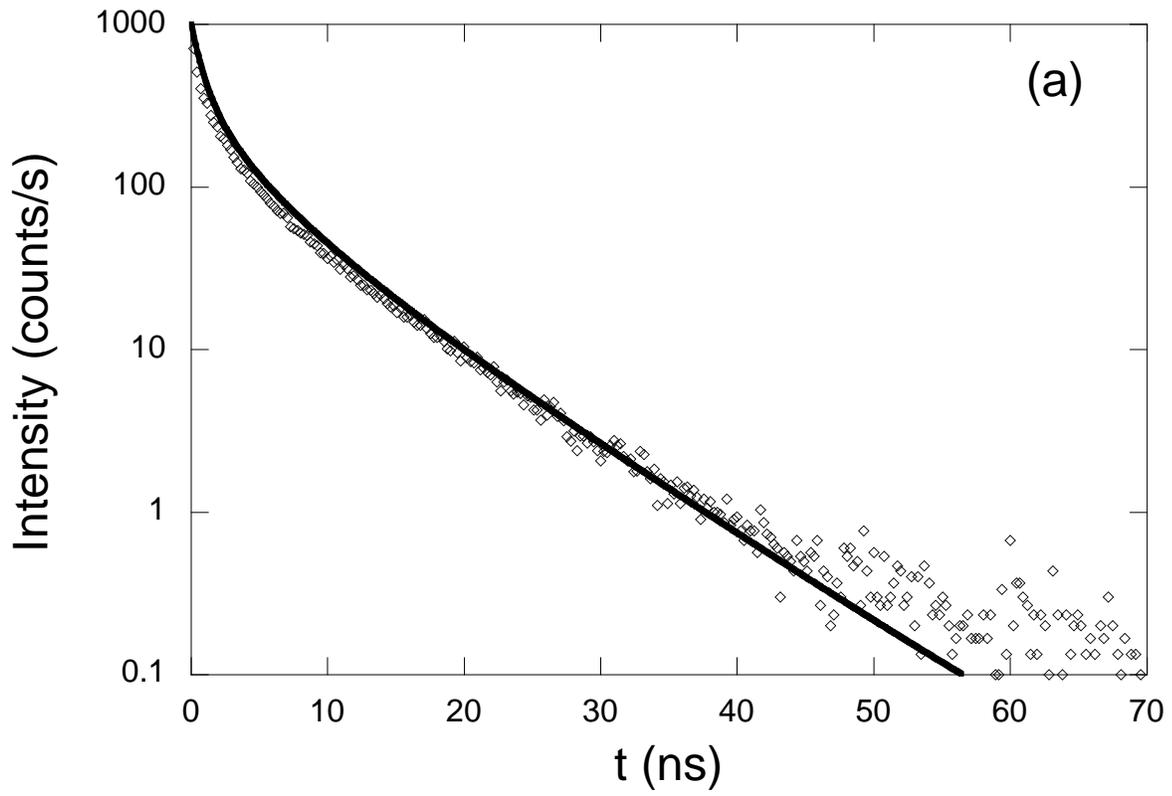
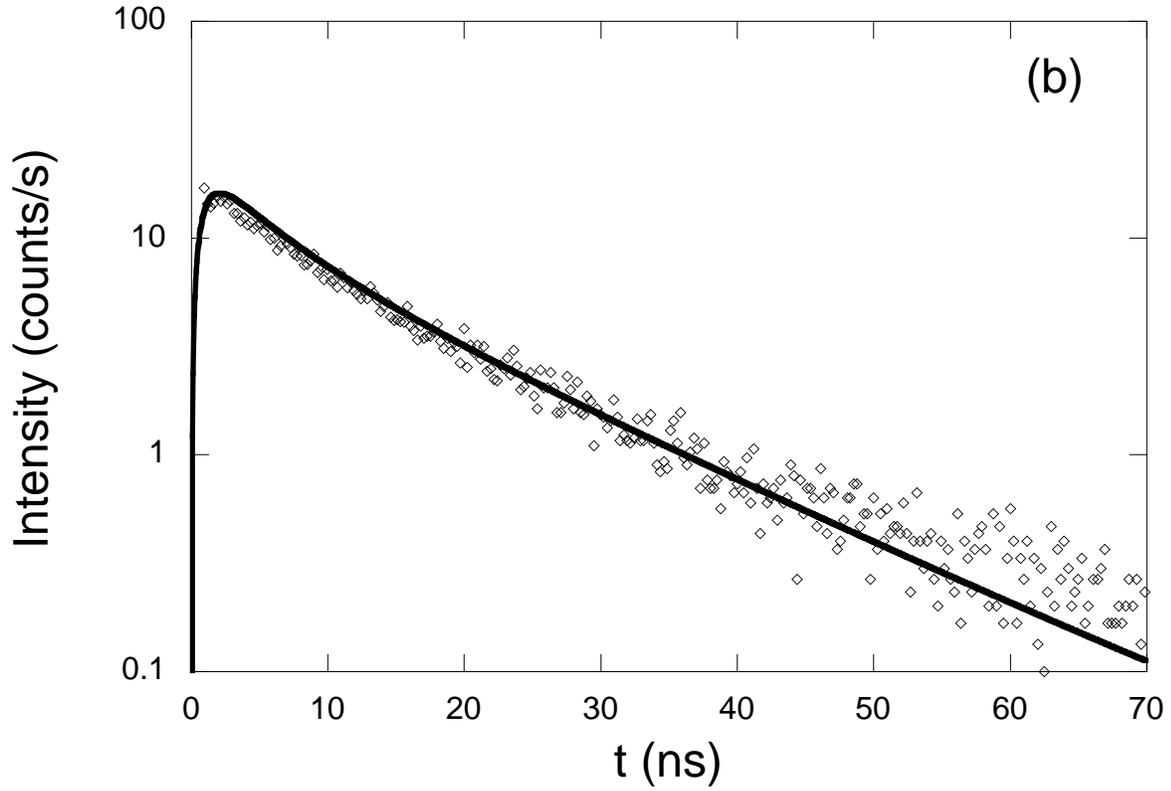
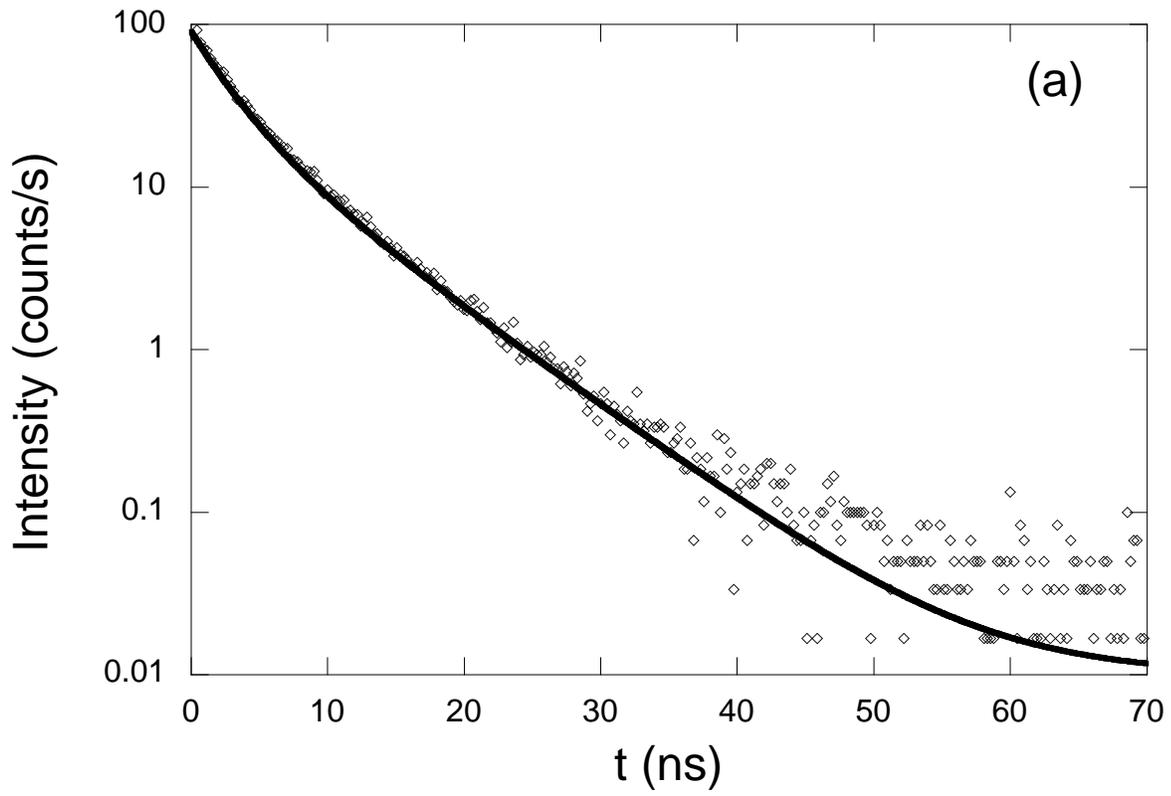
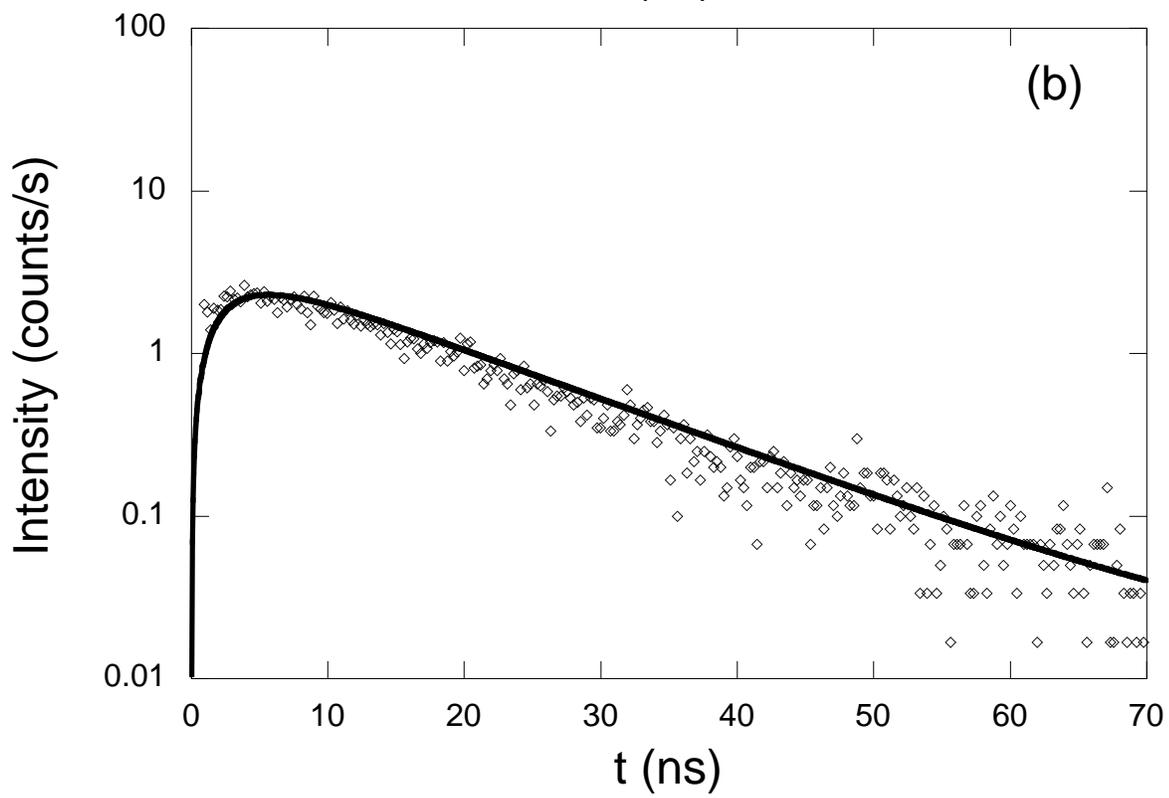
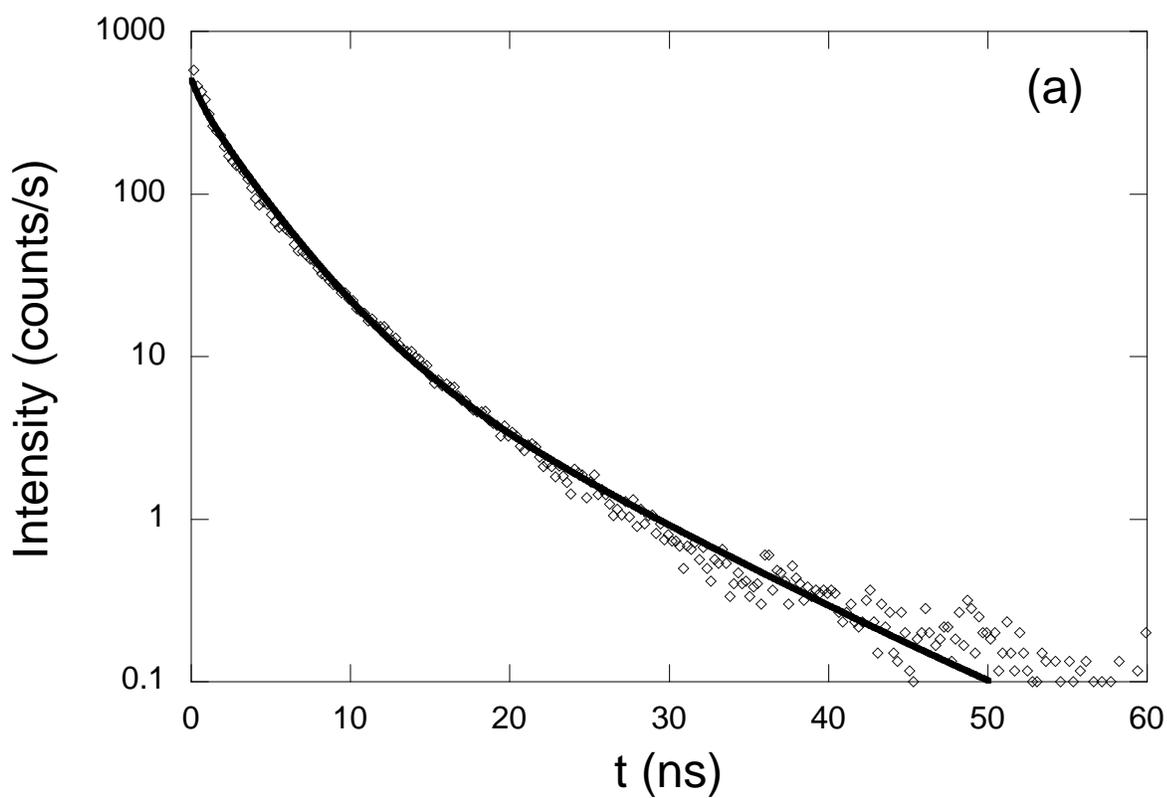
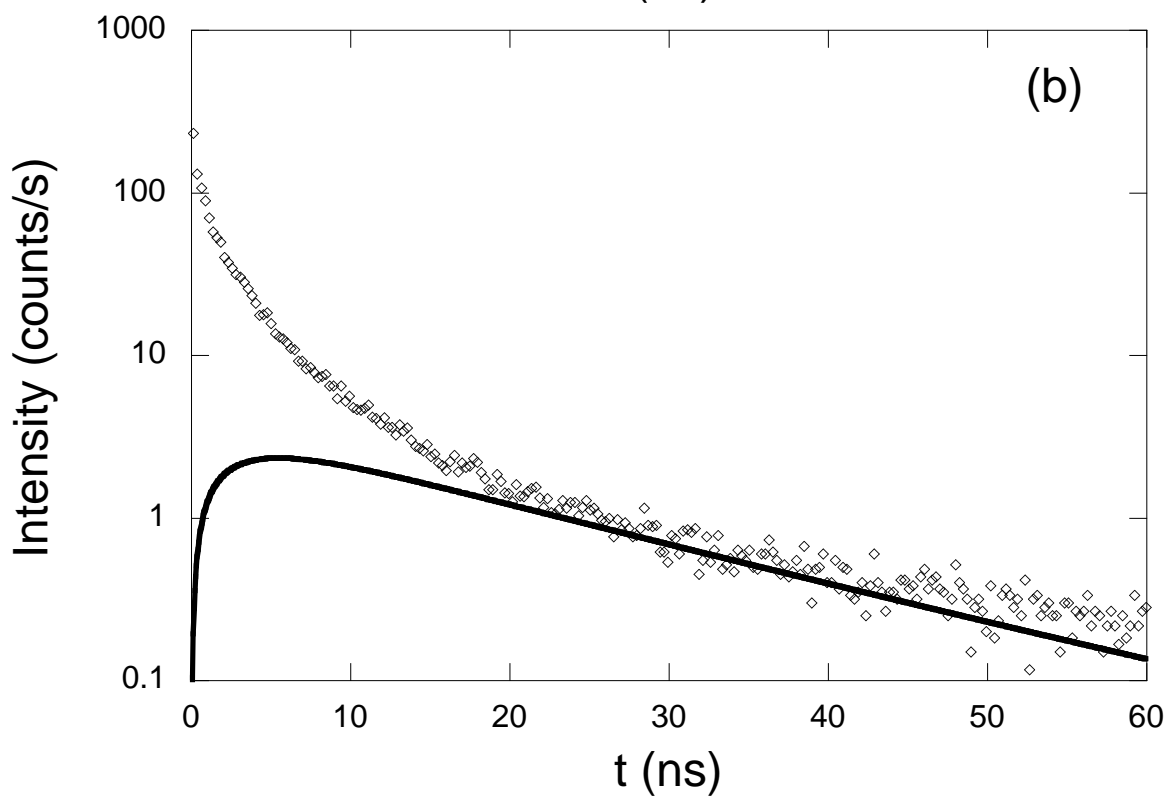
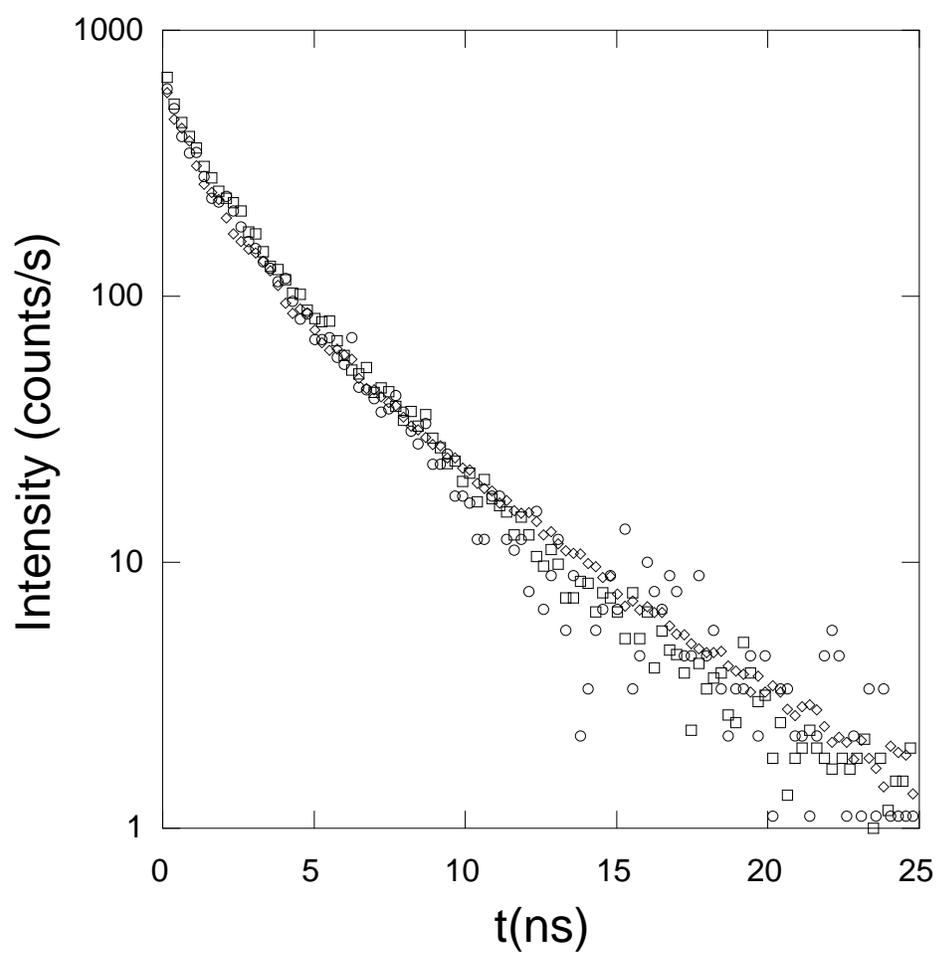
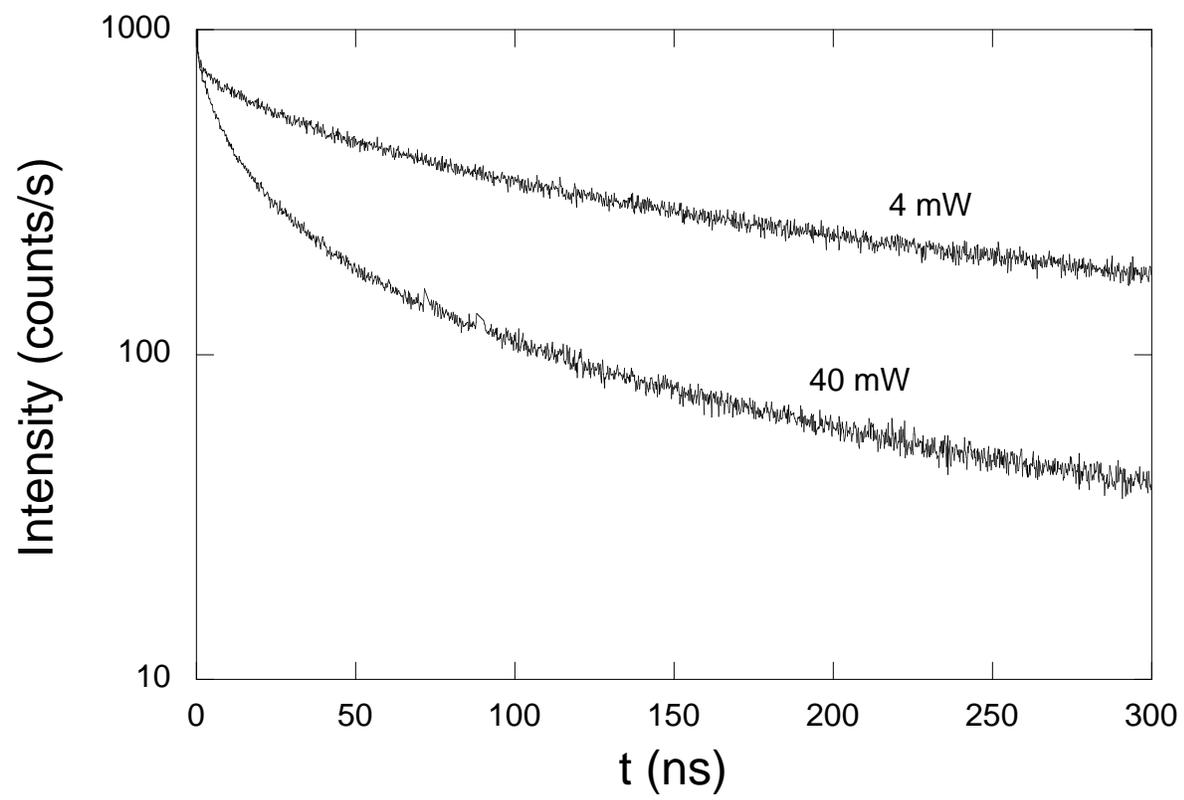
\end{document}